\documentclass[aps,prl,twocolumn,groupedaddress,showpacs]{revtex4}
\usepackage{graphicx}
\begin{document}

\title{Determination of the magnetic anisotropy axes of 
single-molecule magnets}

\author{W. Wernsdorfer$^1$, N. E. Chakov$^2$, and G. Christou$^2$}

\affiliation{
$^1$Lab. L. N\'eel, associ\'e \`a l'UJF, CNRS, BP 166,
38042 Grenoble Cedex 9, France\\
$^2$Dept. of Chemistry, Univ. of Florida,
Gainesville, Florida 32611-7200, USA
}

\date{\today}

\begin{abstract}
Simple methods are presented allowing the determination of
the magnetic anisotropy axes of a crystal 
of a single-molecule magnet (SMM).
These methods are used to determine an upper bound 
of the easy axis tilts in a standard Mn$_{12}-$Ac crystal. 
The values obtained in the present study are significately smaller than
those reported in recent high frequency electron paramagnetic
resonance (HF-EPR) 
studies which suggest distributions of hard-axes tilts.
\end{abstract}

\maketitle

Single-molecule magnets (SMMs) are among the 
smallest nanomagnets that exhibit 
magnetization hysteresis, a classical property of macroscopic 
magnets~\cite{Christou00,Sessoli93b,Sessoli93,Aubin96,Boskovic02}. 
They straddle the interface 
between classical and quantum mechanical behavior because they also 
display quantum tunneling of magnetization
~\cite{Novak95,Friedman96,Thomas96,Sangregorio97,Hill98,Aubin98,Kent00b,Soler04,Tasiopoulos04} 
and quantum phase interference~\cite{Garg93,WW_Science99}. 
These molecules comprise several magnetic ions, 
whose spins are coupled by strong exchange interactions
to give a large effective spin. The molecules are regularly
assembled within large crystals, with all the molecules often having the same
orientation. Hence, macroscopic measurements can give direct access
to single molecule properties. 

An important tool to tune the quantum properties is 
the application of transverse fields.
In particular, the tunnel splitting can be tuned by 
a transverse field via the $S_xH_x$ and $S_yH_y$ Zeeman terms
of the spin Hamiltonian~\cite{Garg93,WW_Science99}.
Therefore, in order to study the tunnel dynamics 
in SMMs, a precise alignment of the field directions is 
necessary. 

\begin{figure}
\begin{center}
\includegraphics[width=.42\textwidth]{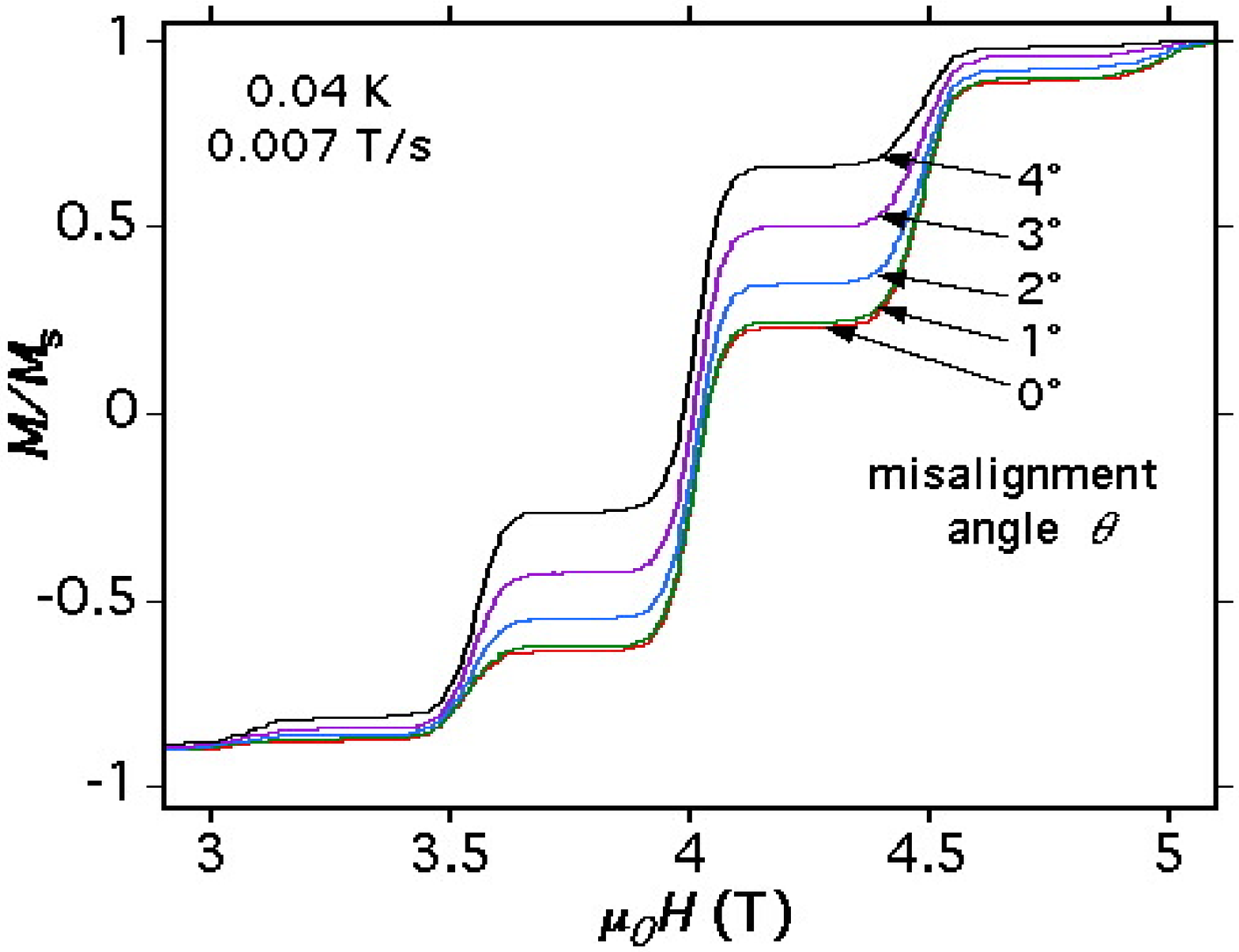}
\includegraphics[width=.42\textwidth]{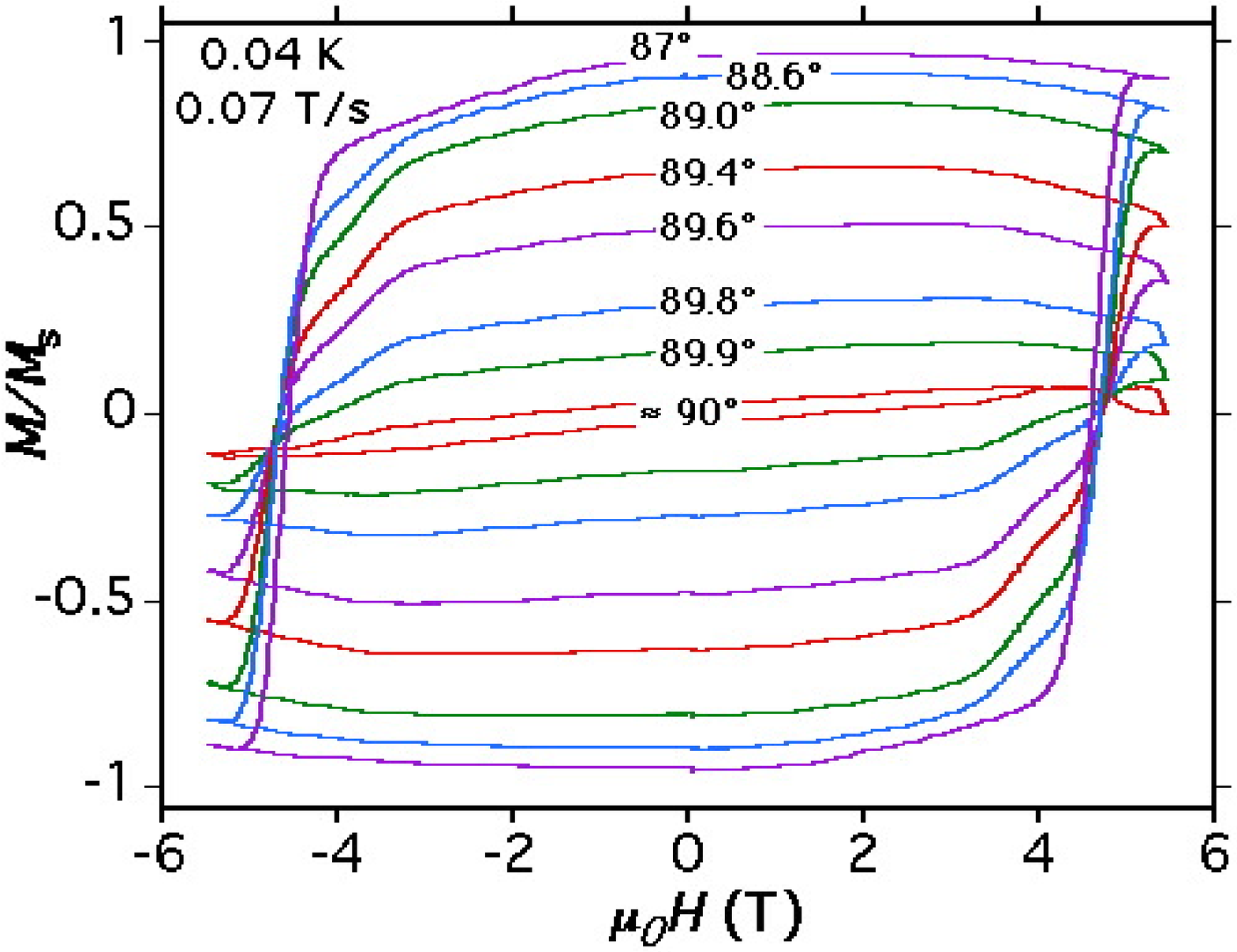}
\caption{(Online color)
(a) Positive part of the hysteresis loop of
a single crystal of Mn$_{12}-$Ac for several 
misalignment angles. The magnetization $M$
along the $c-$axis of the crystal is
normalized by its saturation value $M_{\rm s}$.
The steps are due to
resonant tunneling between the spin ground state
with the quantum number
$m = -10$ and excited state $m =$ 4, 3, \ldots, 0. 
No clear difference is observed between the misalignment
angles of 0 and 1$^{\circ}$.
(b) Similar hysteresis loops but for angles close 
to the hard plane (along the $a-$ axis of the crystal).}
\label{hyst_angle}
\end{center}
\end{figure}

In this communication, we present 
three simple methods to align the magnetic field
that we have used for all our published micro-SQUID 
and micro-Hall probe measurements.
Applied to the standard Mn$_{12}-$Ac SMM, 
these methods have allowed us to
estimate an upper bound of the distribution of easy
axes. We found values that are significately smaller
than those of recent high frequency electron paramagnetic
resonance (HF-EPR) studies~\cite{Hill_PRL03,Hill_0401515,Barco_0404390} 
which suggest distributions of hard-axes tilts with widths of 
1.7$^{\circ}$ and 1.3$^{\circ}$ for standard 
and deuterated Mn$_{12}-$Ac single-molecule magnets.
A distribution of internal transverse magnetic fields 
was also suggested for the Mn$_{12}-$BrAc SMM with hard-axes 
tilts of 7.3$^{\circ}$~\cite{Barco_PRB04}.

All measurements were performed using a 
2D electron gas micro-Hall probe. 
The high sensitivity  allows the study of single 
crystals of SMMs on the order of 10 to 500 $\mu$m. 
The sample of the present study was $20 \times 6 \times 5$ $\mu$m$^3$.
The field can be applied in any direction by separately 
driving three orthogonal coils.
In this study, the fields were rotated in a plane 
given by the $a-$ and $c-$axes 
of a single crystal of Mn$_{12}-$Ac.
The crystal was attached to the Hall probe so that it
measured mainly the magnetization along the $c-$axis of the 
crystal.

\begin{figure}
\begin{center}
\includegraphics[width=.35\textwidth]{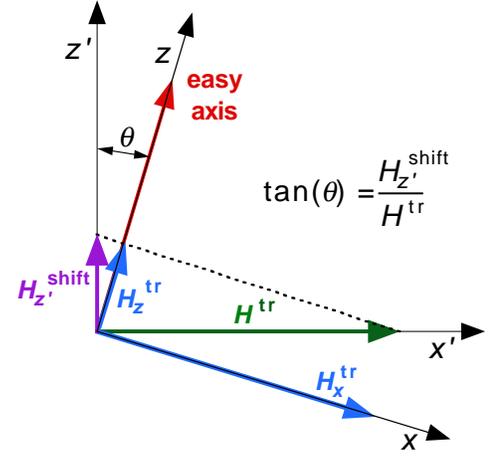}
\caption{(Online color)
Scheme of the coordinate system $(x,z)$ of the 
magnetic anisotropy of a SMM 
where the easy axis of magnetization is along $z$ 
and a coordinate system $(x',z')$.
The latter is rotated by a misalignment angles $\theta$ 
with respect to $(x,z)$. The applied 
field $H$ is swept along $z'$ in the presence of a 
constant transverse field $H^{\rm tr}$ applied along $x'$. 
The latter can be decomposed into $H_x^{\rm tr}$
and $H_z^{\rm tr}$ terms along $x$ and $z$, 
respectively (Fig.~\ref{vectors}) 
}
\label{vectors}
\end{center}
\end{figure}

The first method to find the easy axis of magnetization 
consists in measuring hysteresis loops as a function 
of the angle of the applied field. 
Typical results for angles close 
to the easy axis of magnetization 
are presented in Fig.~\ref{hyst_angle}a 
showing faster relaxation
for larger misalignment angles. This behavior can be understood by
separating the applied field into two components, one parallel
and the other transverse to the easy axis of magnetization.
The transverse component increases in general the tunnel rate
via the $S_xH_x$ and $S_yH_y$ Zeeman terms
of the spin Hamiltonian. Only for special cases, a decrease
of the tunnel rate can be observed that is due
to quantum interference~\cite{Garg93,WW_Science99}.
The positions of the tunnel resonances are only slightly
affected by a small misalignment angle~\cite{remark1}.
This method is therefore not very sensitive.

A second, very similar method consists of measuring
hysteresis loops as a function of angle of 
the applied field, but for angles
close to the hard plane of magnetization.
Typical results are given in Fig.~\ref{hyst_angle}b, 
showing that the hysteresis loop is nearly closed 
when the field is aligned transverse to the easy
axis. This method is more sensitive than the
first one, but is often not very convenient.

\begin{figure}
\begin{center}
\includegraphics[width=.42\textwidth]{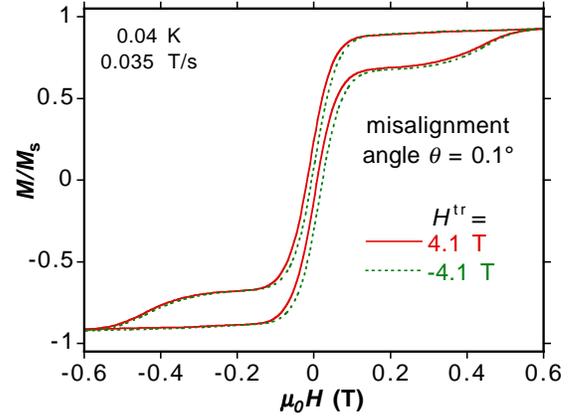}
\caption{(Online color) Normalized magnetization along the $c-$axis
of the crystal versus a field applied at a misalignment angle 
$\theta = 0.1^{\circ}$. A constant transverse field 
$H_{\rm tr} = \pm 4.1$ T is applied leading to clear field shifts 
$H_{z'}^{\rm shift} = H^{\rm tr} \tan(\theta) \approx$ O.OO7 T 
(Fig.~\ref{vectors})
of the zero field resonance.}
\label{hyst_mis}
\end{center}
\end{figure}

A third method is shown in Fig.~\ref{vectors}.
Let's call $(x,y,z)$ the coordinate system of the 
magnetic anisotropy of a SMM 
where the easy axis of magnetization is along $z$. 
Another coordinate system $(x',y',z')$ 
is rotated by two misalignment angles $(\theta,\phi)$ 
with respect to $(x,y,z)$. 
The purpose of the method is to find 
the misalignment angles. For the sake of simplicity, 
the following discussion is in two dimensions. 
A generalization to three dimensions is straightforward 
and is discussed below. 
The two reduced coordinate systems $(x,z)$ and $(x',z')$ 
are misaligned by $\theta$ (Fig.~\ref{vectors}). 
The method consists of sweeping the applied 
field $H$ along $z'$ in the presence of a 
constant transverse field $H^{\rm tr}$ applied along $x'$. 
The latter can be decomposed into $H_x^{\rm tr}$
and $H_z^{\rm tr}$ along $x$ and $z$, 
respectively (Fig.~\ref{vectors}). 

$H_x^{\rm tr}$ modifies the tunnel rates 
of the spin system whereas $H_z^{\rm tr}$ shifts all 
resonance positions by the quantity $H_{z'}^{\rm shift}$
along $z'$ (Fig.~\ref{vectors}):
\begin{equation}
	H_{z'}^{\rm shift} = H^{\rm tr} \tan(\theta).
\label{eq_H_shift}
\end{equation}

Fig.~\ref{hyst_mis} exhibits a typical measurement 
for Mn$_{12}-$Ac for a misalignment angle of $\theta = 0.1^{\circ}$ 
and $H^{\rm tr} = \pm 4.1$ T 
leading to $H_{z'}^{\rm shift} \approx \pm 0.007$ T. 
The latter can be measured easily, thereby allowing 
a field alignment much better than $0.1^{\circ}$.

In order to generalize the above method 
to a three dimensional alignment, it is convenient
to choose two orthogonal planes. Firstly,
the projection of the easy axis 
into one plane is measured. Then, the 
orthogonal plane is rotated so that it contains 
the easy axis projection. Finally,
it is sufficiant to apply again the
above method in this orthogonal plane
in order to find the easy axis. 
The final result can be checked
by sweeping the field along the easy axis
in the presence of a constant transverse
field. No net shifts of the resonance fields
should be observed when comparing both
parts of the hysteresis loops.

It is also important to note that the above
method works in the thermally activated regime
and even above the blocking temperature.
In particular, only small transverse fields
are needed at higher temperatures.
For easy plane anisotropies and more complex
anisotropies, analogues versions
can be figured out easily.

\begin{figure}
\begin{center}
\includegraphics[width=.42\textwidth]{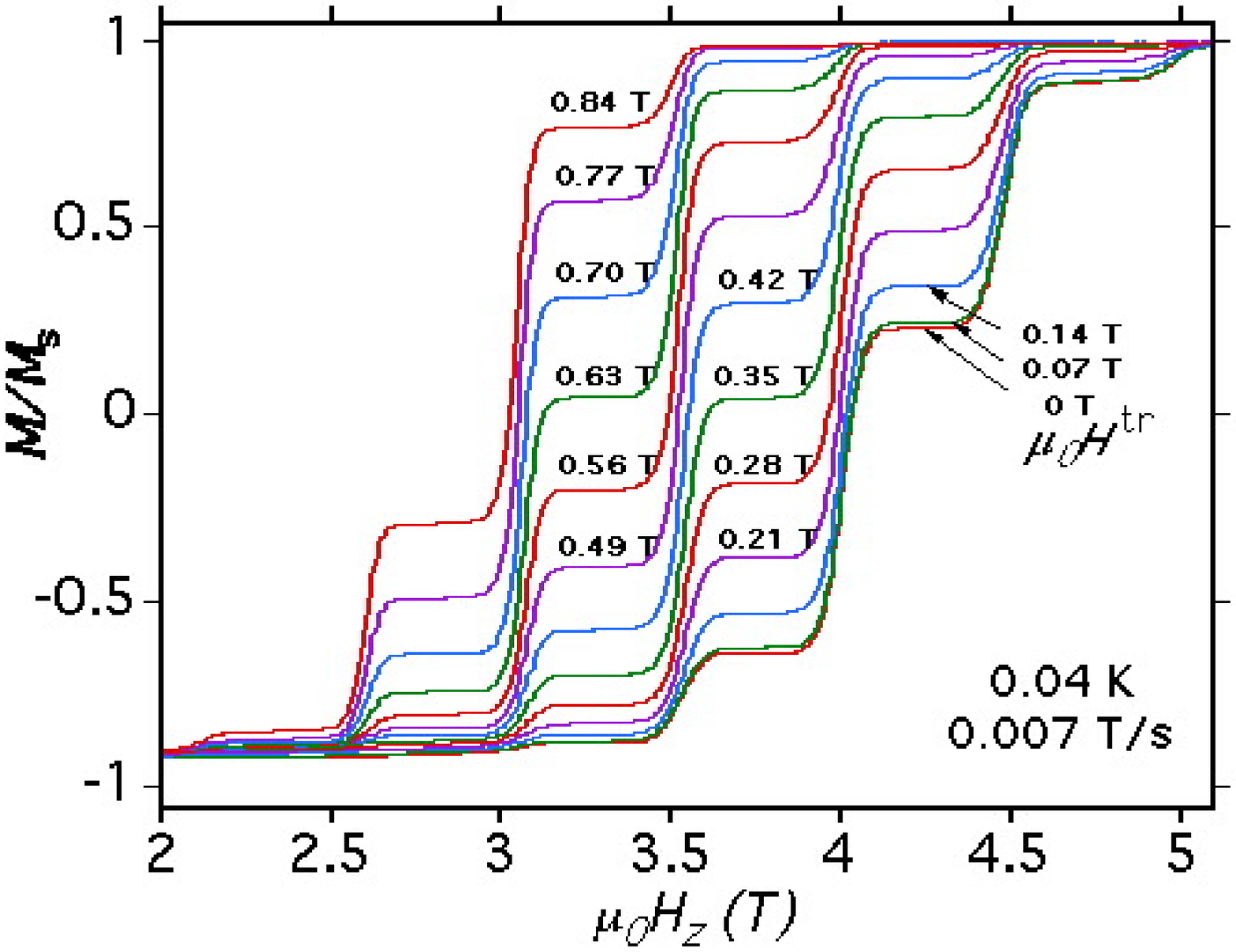}
\includegraphics[width=.42\textwidth]{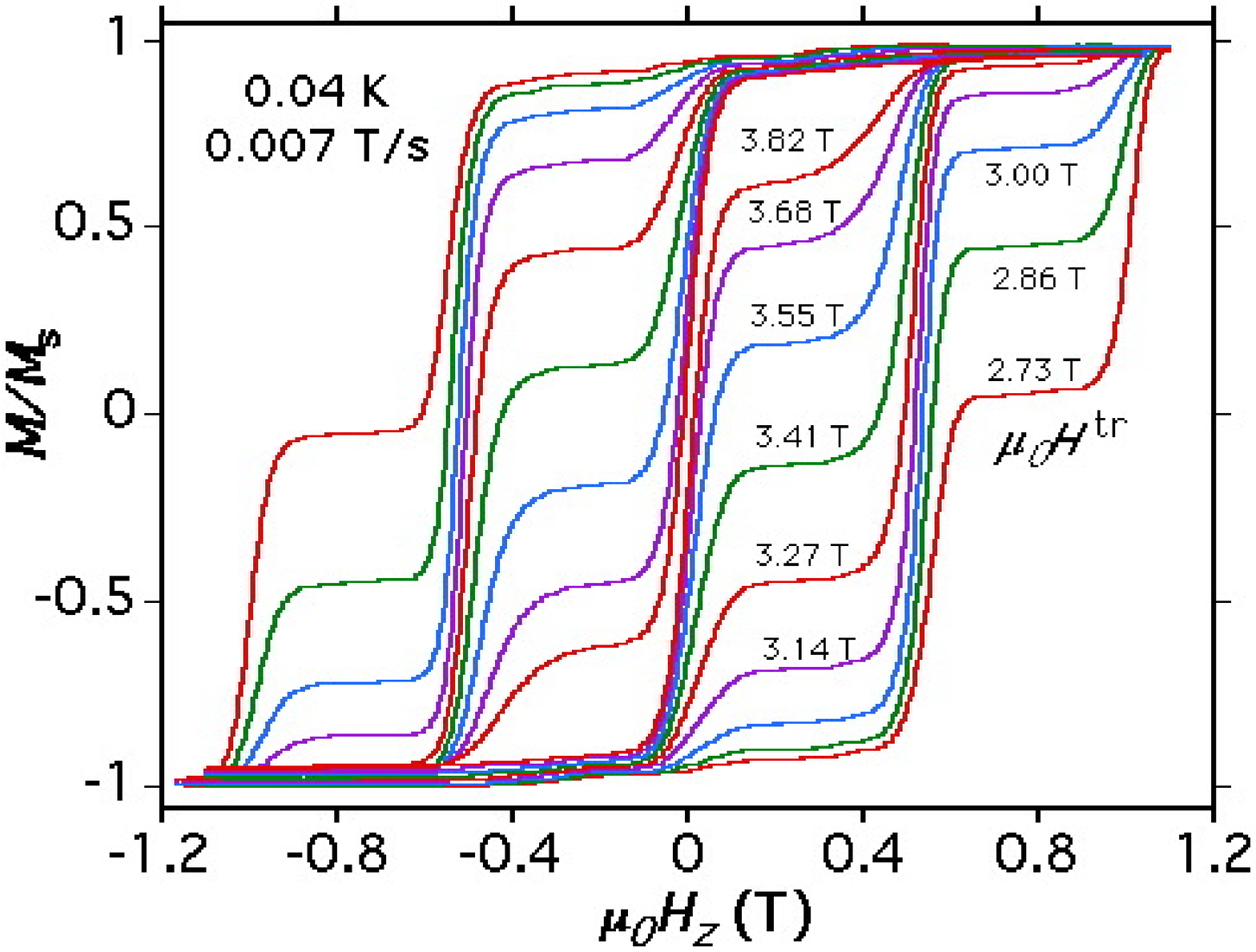}
\caption{(Online color) (a and b) Normalized magnetization along the $c-$axis
of the crystal versus applied field along the $c-$axis 
for several constant transverse fields $H_{\rm tr}$.  
Although the transverse fields increase the tunnel rates, no 
significant broadening of the resonance fields is observed.}
\label{hyst_tr}
\end{center}
\end{figure}

\begin{figure}
\begin{center}
\includegraphics[width=.42\textwidth]{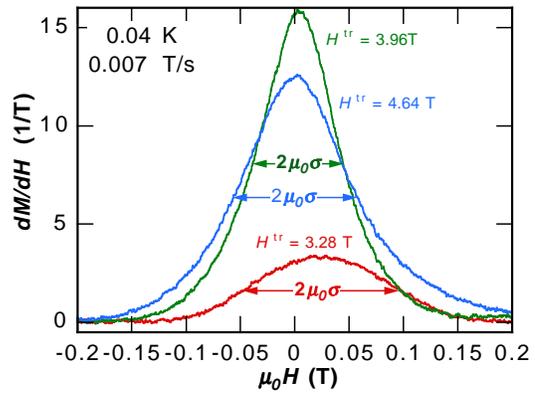}
\caption{(Online color) Derivative $dM/dH$ of three hysteresis
curves at the zero field resonance for a field sweep from negative
to positive fields. The definition of the width $\sigma$
at the half-width-at-half-maximum is shown.}
\label{derivative}
\end{center}
\end{figure}

\begin{figure}
\begin{center}
\includegraphics[width=.42\textwidth]{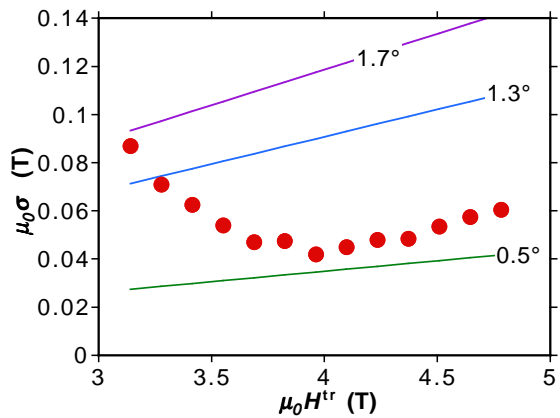}
\caption{(Online color) Half-width-at-half-maximum $\sigma$
versus transverse field for the zero field resonance.}
\label{sigma}
\end{center}
\end{figure}

We use here our methods to determine an upper bound 
of the easy axis tilts in a standard Mn$_{12}-$Ac crystal. 
We first align our fields 
with respect to the easy axis of Mn$_{12}-$Ac
using the above methods. 
We measure then all tunnel transitions as a 
function of transverse field (Fig.~\ref{hyst_tr}) and study 
their widths $\sigma$. 
Fig.~\ref{derivative} presents the first derivative of the 
magnetization $dM/dH$ for the zero field resonance 
for several transverse fields. We defined the resonance width $\sigma$
as the half-width-at-half-maximum, in accordance with Ref. 
\cite{Hill_PRL03,Hill_0401515,Barco_0404390}.
Fig.~\ref{sigma} presents $\sigma$
as a function of a transverse field showing a minimum
of the width at about 4 T.
Recent HF-EPR studies~\cite{Hill_PRL03,Hill_0401515,Barco_0404390} 
suggested that there 
are distributions of hard-axes tilts with widths of 
1.7$^{\circ}$ and 1.3$^{\circ}$ for standard  
and deuterated Mn$_{12}-$Ac, respectively. Fig.~\ref{sigma} shows 
the expected width of the zero field resonance
supposing that it is only due to the distribution of hard-axes tilts.
Our results suggest an upper bound of 0.5$^{\circ}$.
The actual hard-axes tilts might be much smaller
because we expect a dipolar broadening of the resonance
lines of about 0.03 T. In addition, higher order
tunneling transitions induced by dipolar and small
superexchange interactions might further broaden
the resonance transition~\cite{WW_PRL02}. We therefore believe that
the hard-axes tilts should not exceed about 0.2$^{\circ}$.

This result is confirmed by the second method. An upper
bound for the hard-axis tilts is given by the angle needed
to get a hysteresis loop that reaches $|M/M_{\rm s}| = 0.5$.
We find $\theta = 89.6^{\circ}$ (Fig.~\ref{hyst_angle}b), 
that is an upper bound
of $0.4^{\circ}$ for the hard-axis tilts.
The actual value might be much smaller because
we neglect here multi-body tunnel effects~\cite{WW_PRL02}
that should be rather strong due to the high transverse fields.

We also applied our methods to the Mn$_{12}-$BrAc SMM 
and could not confirm the hard-axes 
tilts of 7.3$^{\circ}$ suggested by 
del Barco et al.~\cite{Barco_PRB04}.
Our results showed that the hard-axes 
tilts in Mn$_{12}-$BrAc might be even smaller
than in Mn$_{12}-$Ac.

Finally, we speculate about the origin of the line widths
observed in the EPR studies~\cite{Hill_PRL03,Hill_0401515,Barco_0404390}. 
We suggest that the observed fine
structures are due to the presence of fast relaxing species
~\cite{remark2} having a smaller magnetic anisotropy
and are tilted with respect to the $c-$axis by about 
$10^{\circ}$~\cite{WW_EPL99}. These species are coupled
via dipolar interactions to the normal ones leading to
multi-body effects (cross relaxations)~\cite{WW_PRL02} 
that might broaden the lines. Such an interpretation
is supported by the fact that Mn$_{12}-$BrAc does not
show anormalous EPR line width broadening~\cite{Petukhov}
because it hardly has fast relaxing species.

In conclusion, we have presented three methods 
that allow the determination of
the magnetic anisotropy axes of a crystal 
of a single-molecule magnet (SMM). 
The precise field alignments are necessary 
when studying quantitatively resonant tunneling
of magnetizations in spin systems like SMMs.

\begin{thebibliography}{10}

\bibitem{Christou00}
G. Christou, D. Gatteschi, D.N. Hendrickson, and R. Sessoli, MRS Bulletin {\bf
  25},  66  (2000).

\bibitem{Sessoli93b}
R. Sessoli, H.-L. Tsai, A.~R. Schake, S. Wang, J.~B. Vincent, K. Folting, D.
  Gatteschi, G. Christou, and D.~N. Hendrickson, J. Am. Chem. Soc. {\bf 115},
  1804  (1993).

\bibitem{Sessoli93}
R. Sessoli, D. Gatteschi, A. Caneschi, and M.~A. Novak, Nature {\bf 365},  141
  (1993).

\bibitem{Aubin96}
S.~M.~J. Aubin and {\it et al.}, J. Am. Chem. Soc. {\bf 118},  7746  (1996).

\bibitem{Boskovic02}
C. Boskovic and et~al., J. Am. Chem. Soc. {\bf 124},  3725  (2002).

\bibitem{Novak95}
M.A. Novak and R. Sessoli,  in {\em Quantum Tunneling of Magnetization-QTM'94},
  Vol.~301 of {\em NATO ASI Series E: Applied Sciences}, edited by L. Gunther
  and B. Barbara (Kluwer Academic Publishers, London, 1995), pp.\ 171--188.

\bibitem{Friedman96}
J.~R. Friedman, M.~P. Sarachik, J. Tejada, and R. Ziolo, Phys. Rev. Lett. {\bf
  76},  3830  (1996).

\bibitem{Thomas96}
L. Thomas, F. Lionti, R. Ballou, D. Gatteschi, R. Sessoli, and B. Barbara,
  Nature (London) {\bf 383},  145  (1996).

\bibitem{Sangregorio97}
C. Sangregorio, T. Ohm, C. Paulsen, R. Sessoli, and D. Gatteschi, Phys. Rev.
  Lett. {\bf 78},  4645  (1997).

\bibitem{Hill98}
S. Hill, J.A.A.J. Perenboom, N.S. Dalal, T. Hathaway, T. Stalcup, and J.S.
  Brooks, Phys. Rev. Lett. {\bf 80},  2453  (1998).

\bibitem{Aubin98}
S.~M.~J. Aubin, N.~R. Dilley, M.~B. Wemple, G. Christou, and D.~N. Hendrickson,
  J. Am. Chem. Soc. {\bf 120},  839  (1998).

\bibitem{Kent00b}
L. Bokacheva, A.D. Kent, and M.A. Walters, Phys. Rev. Lett. {\bf 85},  4803
  (2000).

\bibitem{Soler04}
M. Soler, W. Wernsdorfer, K. Folting, M. Pink, and G. Christou, J. Am. Chem.
  Soc.  2156  (2004).

\bibitem{Tasiopoulos04}
A.J. Tasiopoulos, A. Vinslava, W. Wernsdorfer, K.A. Abboud, and G. Christou,
  Angew. Chem. Int. Ed. Engl. {\bf 43},  2117  (2004).

\bibitem{Garg93}
A. Garg, EuroPhys. Lett. {\bf 22},  205  (1993).

\bibitem{WW_Science99}
W.Wernsdorfer and R. Sessoli, Science {\bf 284},  133  (1999).

\bibitem{Hill_PRL03}
S. Hill, R.~S. Edwards, S.~I. Jones, N.~S. Dalal, and J.~M. North, Phys. Rev.
  Lett. {\bf 90},  217204  (2003).

\bibitem{Hill_0401515}
S. Hill, S. Takahashi, R.~S. Edwards, J.~M. North, and N.~S. Dalal,
  cond-mat/0401515  .

\bibitem{Barco_0404390}
E. del Barco, A.~D. Kent, S. Hill, J.~M. North, N.~S. Dalal, E.~M. Rumberger,
  D.~N. Hendrickson, N. Chakov, and G. Christou, cond-mat/0404390  .

\bibitem{Barco_PRB04}
E. del Barco, A.~D. Kent, N.~E. Chakov, L.~N. Zakharov, A.~L. Rheingold, D.~N.
  Hendrickson, and G. Christou, Phys. Rev. B {\bf 69},  20411  (2004).

\bibitem{remark1}  
The position of the resonances should increase roughly by the factor of 
$\cos(\theta)^{-1} \approx 1$ for small angles $\theta$.

\bibitem{WW_PRL02}
W. Wernsdorfer, S. Bhaduri, R. Tiron, D.~N. Hendrickson, and G. Christou, Phys.
  Rev. Lett. {\bf 89},  197201  (2002).

\bibitem{WW_EPL99}
W. Wernsdorfer, R. Sessoli, and D. Gatteschi, EuroPhys. Lett. {\bf 47},  254
  (1999).

\bibitem{Ruiz98}
D. Ruiz, Z. Sun, B. Albela, K. Folting, J. Ribas, G. Christou, and D.~N.
  Hendrickson, Angew. Chem. Int. Ed. Engl. {\bf 37},  300  (1998).

\bibitem{Sun98}
Z. Sun, D. Ruiz, E. Rumberger, C.~D. Incarvito, K. Folting, A.~L. Rheingold, G.
  Christou, and D.~N. Hendrickson, Inorg. Chem. {\bf 37},  4758  (1998).

\bibitem{Sun99}
Z. Sun, D. Ruiz, N.~R. Dilley, M. Soler, J. Ribas, K. Folting, B. Maple, G.
  Christou, and D.~N. Hendrickson, Chem. Commun.  1973  (1999).

\bibitem{Aubin01}
S.~M.~J. Aubin, Z. Sun, H.~J. Eppley, E.~M. Rumberger, I.~A. Guzei, K. Folting,
  P.~K. Gantzel, A.~L. Rheingold, G. Christou, and D.~N. Hendrickson, Inorg.
  Chem. {\bf 40},  2127  (2001).

\bibitem{Aubin01b}
S.~M.~J. Aubin, Z. Sun, H.~J. Eppley, E.~M. Rumberger, I.~A. Guzei, K. Folting,
  P.~K. Gantzel, A.~L. Rheingold, G. Christou, and D.~N. Hendrickson,
  Polyhedron {\bf 20},  1139  (2001).

\bibitem{Soler03}
M. Soler, W. Wernsdorfer, Z. Sun, D. Ruiz, J.~C. Huffman, D.~N. Hendrickson,
  and G. Christou, Polyhedron {\bf 22},  1783  (2003).

\bibitem{Soler03b}
M. Soler, W. Wernsdorfer, Z. Sun, J.~C. Huffman, D.~N. Hendrickson, and G.
  Christou, Chem. Commun.  2672  (2003).

\bibitem{remark2}  
Several authors have pointed out that in the Mn$_{12}$ carboxylate family
different isomeric forms give rise to different relaxation rates.
This was first observed in Mn$_{12}$-ac 
\cite{Ruiz98,Aubin98,Sun98,Sun99}
and has been studied in detail \cite{Aubin01,Aubin01b,Soler03,Soler03b}. 
We found that a minor species of Mn$_{12}$-ac \cite{WW_EPL99}, randomly
distributed in the crystal, exhibits a faster
relaxation rate which becomes temperature independent below 0.3 K. 


\bibitem{Petukhov}
K. Petukhov, S. Hill, N. E. Chakov, G. Christou, and K. Abboud,
cond-mat/0403435.
       
\end{thebibliography}

\end{document}